\documentclass[twocolumn, a4paper,pre]{revtex4}

\usepackage{float}
\usepackage{psfrag}
\usepackage{graphicx}
\usepackage{amsmath}

\newcommand{\nn}{{\mathbf{n}}}
\newcommand{\qq}{{\mathbf{q}}}
\newcommand{\msd}{\langle \delta u_\nn(t)^2\rangle}
\newcommand{\uun}{\mathbf u (\mathbf n,t)}
\newcommand{\uuq}{\mathbf u (\mathbf q, \omega)}

\newcommand{\boldeta}{\boldsymbol w (\mathbf n,t)}
\newcommand{\boldetaq}{\boldsymbol w (\mathbf q,\omega)}

\newcommand {\be} {\begin{equation}}
\newcommand {\ee} {\end{equation}}
\newcommand {\bea} {\begin{eqnarray}}
\newcommand {\eea} {\end{eqnarray}}

\begin{document}

\title{Tracer Particles in Two--Dimensional Elastic Networks Diffuse Logarithmically Slow 
}

\author{Ludvig Lizana}
\email{ludvig.lizana@umu.se}
\affiliation{Integrated Science Lab, Department of Physics, Ume{\aa}
University, SE--901 87 Ume{\aa}, Sweden}

\author{Tobias Ambj\"ornsson}
\affiliation{Department of Astronomy and Theoretical Physics, Lund University,
  S\"olvegatan 14A, SE--223 62 Lund, Sweden}
 
\author{Michael A. Lomholt}
\affiliation{MEMPHYS, Department of Physics, Chemistry and Pharmacy,
University of Southern Denmark, DK-5230 Odense M, Denmark}

\date{\today}



\begin{abstract}
Several experiments on tagged molecules or particles in living systems suggest that they move anomalously slow -- their mean squared displacement (MSD) increase slower than linearly with time. Leading models aimed at understanding these experiments predict that the MSD grows as a power law with a growth exponent that is smaller than unity. However, in some experiments the exponent is so small $(\sim 0.1-0.2)$ that they hint towards other mechanisms at play. In this paper, we theoretically demonstrate how in-plane collective modes excited by thermal fluctuations in a two dimensional membrane lead to logarithmic time dependence for the mean square displacement of a tracer particle.

\end{abstract}

\maketitle


\section{Introduction}

Macromolecules in crowded environments such as living cells often diffuse slower than if they were free. In free unbiased diffusion the  mean squared displacement (MSD) traveled by the macromolecule $\langle r^2(t)\rangle$ would be proportional to time $t$.  However, several single--particle tracking experiments  show that $\langle r^2(t)\rangle/t$ is a decaying function of time \cite{barkai2012single}. This behaviour is referred to as anomalous and  has been observed in tracking experiments such as fluorescently labelled  mRNA molecules  in the  intracellular volume \cite{golding2006physical},  lipid granules in membranes of yeast cells \cite{jeon2011vivo},  membrane proteins in human kidney cells \cite{weigel2011ergodic} and in crowded media \cite{szymanski2009}. Altogether this suggest that anomalous dynamics is a natural part of  living systems.

To properly model and understand  tracking experiments in cells, we need better understanding of the origin of anomalous dynamics. As of today  there are  three leading physical models: fractional Brownian motion (FBM), continuous time random walk (CTRW), and diffusion on fractals. FBM is a generalised Gaussian process with slowly decaying correlations \cite{mandelbrot1968fractional}. CTRW describes a  particle that jumps in random directions with the time between consecutive jumps drawn from a waiting time distribution (possibly with a diverging first moment)\cite{metzler2000random}. For diffusion on fractals the fractal is sometimes considered to be a percolation cluster \cite{mardoukhi2015geometry}. 

These frameworks share two important features. (i) They describe single--particle  diffusion processes, and (ii) they give a MSD that grows as a sub--linear power law, that is $\langle r^2(t)\rangle/t \propto t^{\alpha-1}$ for some exponent $\alpha <1$.  Due to (ii), the common approach to analyse tracer particle motion is therefore to fit the MSD with a power--law. It is, however, not always clear that a power--law is the best functional form to use in the fitting procedure.  

To better understand what other types of functional forms that may be useful for MSD fitting and  how a system's collective modes can influence the motion of a single particle, we here investigate a simple physical system where the MSD of a tracer particle does {\em not} follow a power--law. Namely, the in-plane dynamics of a tracer particle in a two dimensional elastic network or solid membrane, where the dissipation happens either due to friction against a solid support or interaction with a surrounding viscous fluid. We quantify how a labeled particle in the membrane (tracer particle) fluctuates within the membrane's plane, i.e., lateral motion, and find  that the MSD grows logarithmically with time.


\section{Elastic network model}
\begin{figure}

\includegraphics[width=0.8\columnwidth]{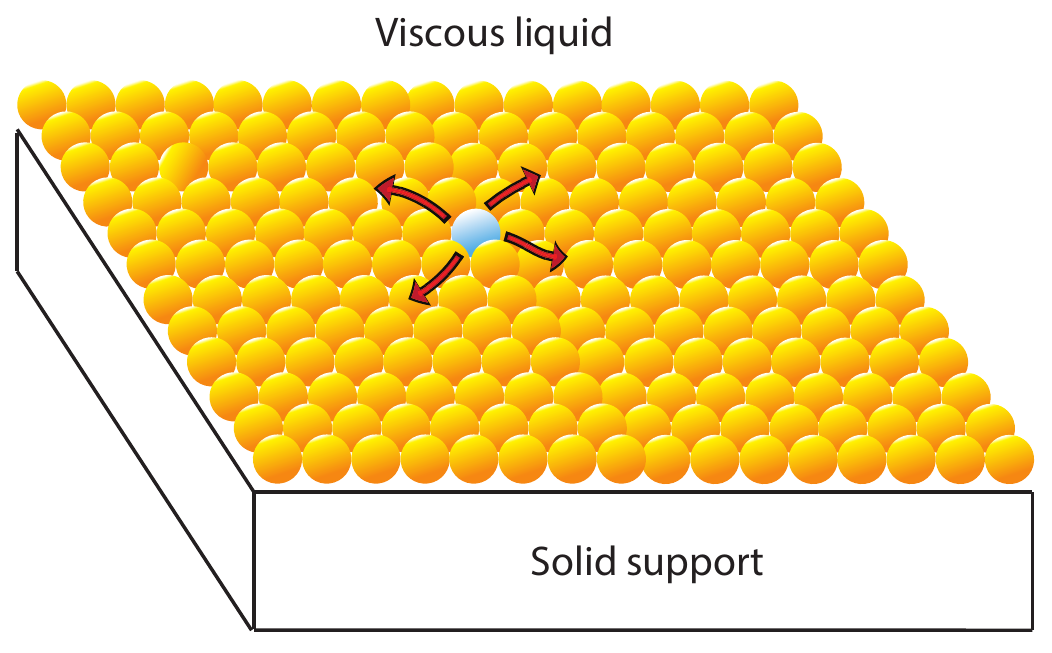}
\caption{Schematic figure of the elastic model. 
We model the membrane as a two dimensional isotropic elastic network with Hookian elastic properties towards both compressional and shear stresses. The motion of each bead is governed by spring forces as well as stochastic and drag forces from the surroundings. We study the collective diffusive motion of a tagged bead (tracer particle) in the membrane (blue). The membrane rests on a solid support with a viscous liquid on the other side.}
\label{fig:membrane_diffusion}
\end{figure}

We model the membrane as a two dimensional isotropic elastic network with beads interacting as a Hooke's law material. Apart from elastic forces, each bead experience stochastic forces from thermal fluctuations and drag forces from the media that surrounds the membrane. We will let the membrane interact with a solid surface on one side, leading to friction with membrane--wall friction constant $\xi$, and a viscous fluid (viscosity $\eta$) on the other side, leading to hydrodynamic interactions between different parts of the membrane (see Fig. \ref{fig:membrane_diffusion}). To transform this setup to a membrane with fluid on both sides, one simply set $\xi=0$ and double the viscosity $\eta\rightarrow 2\eta$ in all equations below.

We label each bead by $\nn=(n_x,n_y)$ and denote $\uun = (u_x({\mathbf n},t),u_y({\mathbf n},t))$ as the  displacement vector with respect to the particles' equilibrium positions.  We will consider length scales that are large compared with the nearest neighbor particle distances such that we can treat  $\nn$ as continuous.
Also, we will assume that the system is large compared with the displacement of the beads during the time of observation so that we can treat the system as infinitely large. Thus we have the equation of motion
\bea \label{eq:elastic_model}
&&\int_{-\infty}^ \infty  d^2n' \, {\cal B}  (\mathbf n- \mathbf n') \cdot \frac{\partial \mathbf u(\mathbf n',t)}{\partial t}  = 
 \\
&&\mu \nabla^2_n \uun + (\lambda+\mu) \nabla_n(\nabla_n\cdot\uun) + \boldeta
\nonumber
\eea
where $\nabla_n=(\frac{\partial}{\partial n_x}, \frac{\partial}{\partial n_y})$, and 
$\lambda$ and $\mu$ are the first and second Lam{\'e} parameters that describe the membrane's elastic properties \cite{landau86,nelson02}. The tensor ${\cal B}  (\mathbf n- \mathbf n')$ represents the friction (the expression for it is given in Fourier space below), and $\boldeta$ is a zero-mean, $\langle \boldeta\rangle=0$, Gaussian white noise. The correlations of $\boldeta$ is determined by the fluctuation-dissipation theorem to be:
\be
\langle w_\alpha (\nn,t) w_\beta (\nn', t') \rangle = 
2 k_BT\delta(t-t') B_{\alpha\beta}(\nn-\nn'),
\ee
where $\langle ... \rangle$ denotes ensemble average and $k_BT$ is thermal energy.

To proceed we write the displacement in terms of Fourier modes as
\be\label{eq:elastic_model_q}
  \uun =\int_{-\infty}^\infty  \frac{d^2q}{(2\pi)^2}
                 \int_{-\infty}^\infty \frac{d\omega}{2\pi} \, 
                 \uuq\, e^{i\qq\cdot \nn
                 -i\omega t}
\ee
This transforms Eq. (\ref{eq:elastic_model}) into
\bea\label{eq:u_q}
&&-i\omega{\cal B}(\qq)\cdot \uuq = 
\\\nonumber 
&&\ \ -\mu q^2  \uuq - (\lambda+\mu) \qq [\qq \cdot  \uuq] + \boldetaq
\eea
where $q = |\mathbf q|$. We point out that an equation similar to Eq. (\ref{eq:elastic_model}) was introduced in \cite{taloni2010correlations} for out--of--plane membrane fluctuations.  In contrast, our equations describe in--plane fluctuations, but it is in any case interesting to note that Eq. (\ref{eq:elastic_model}) overlaps with the equations in \cite{taloni2010correlations} if we set $\lambda+\mu=0$.

In order to solve Eq. (\ref{eq:u_q}), it is convenient to split the displacement vector and friction tensor  into in--plane longitudinal ($l$) and in--plane transverse ($t$) directions 
\be
\uuq =  u^l(\qq,\omega) \hat{\mathbf q }+ u^t(\qq,\omega) \hat{ \mathbf  t }
\ee
\be
{\cal B}(\qq) = B^l(q) \hat{\mathbf q } \hat{\mathbf q } + B^t(q) \hat{\mathbf t } \hat{\mathbf t }
\ee
where $\hat{\mathbf t}$ is orthogonal to $\hat{\mathbf q}={\mathbf q}/q$, and \cite{lomholt2006mechanics}
\begin{align}
B^l(q)&=\xi + 2\eta q\\
B^t(q)&=\xi + \eta q
\end{align}
After splitting, we find from Eq. (\ref{eq:elastic_model_q}) that
\be\label{eq:u_long}
u^l(\qq,\omega) =\frac{w^l(\qq,\omega)}{q^2 (\lambda+2\mu)  - i\omega B^l(q) }
\ee
\be \label{eq:u_trans}
u^t(\qq,\omega) =\frac{w^t(\qq,\omega)}{q^2 \mu  - i\omega B^t(q)}
\ee

Equations (\ref{eq:u_long}) and (\ref{eq:u_trans}) are our starting point for further analysis. We will consider a tracer particle that is pushed around by the in--plane noise. To characterise its motion, we will calculate the tracer particle's MSD.

\section{In--plane tracer particle fluctuations in a membrane}

To analyse a tracer particle's motion in a fluctuating membrane, we study how a tagged piece of the membrane  (i.e a bead in the elastic network) explores space. In particular we calculate the MSD $\msd \equiv \langle[\uun - \mathbf{u}(\nn,0)]^2\rangle$ in the membrane's plane. 
We will assume that the system is in a stationary state, which means that $\langle \mathbf{u}(\nn,t)\cdot\mathbf{u}(\nn,t') \rangle=\langle \mathbf{u}(\nn,|t-t'|)\cdot\mathbf{u}(\nn,0) \rangle$, and leads to
\be\label{eq:defMSD}
\msd= 2 [\langle \mathbf{u}(\nn,0)\cdot\mathbf{u}(\nn,0) \rangle - \langle \mathbf{u}(\nn,t) \cdot \mathbf{u}(\nn,0) \rangle].
\ee
As before we split the correlation $\langle \uun \cdot \mathbf u (\mathbf n,t')\rangle$ into its longitudinal part, $C^l(t)=\langle u^l(t) u^l(0)\rangle$, and corresponding transverse part:
\be
\langle \uun \cdot \mathbf u (\mathbf n,t')\rangle = C^l(t-t') + C^t(t-t'),
\ee
where in terms of $C^t(t)$ and $C^l(t)$ the $\msd$ is
\be\label{eq:msd_c}
\msd =2 \left([C^t(0)-C^t(t)] + [C^l(0)-C^l(t)]\right).
\ee

First we calculate $C^t(t)$.  To achieve this,  we use $u^t(\qq,\omega)$ from Eq. (\ref{eq:u_trans})  and  the noise autocorrelation function 
\be
\langle w_\alpha (\qq, \omega) w_\beta (\qq',\omega') \rangle = 
 2 k_BT (2\pi)^{3}  \delta(\omega+\omega') \delta(\qq+\mathbf q') B_{\alpha\beta}(\qq).
\ee
This leads to
\bea
&&C^t(|t-t'|) = 
		\int_{-\infty}^\infty  \frac{d^2 q}{(2\pi)^2}
                \int_{-\infty}^\infty \frac{d\omega}{2\pi} \,
                e^{-i\omega (t-t')} 
                \nonumber\\
                 &&\times\frac{2k_BT}{B^t(q)}
                 \frac{1}{[\mu q^2/B^t(q)]^2 + \omega^2}
                 \\
&&=				 \nonumber
				 \frac{k_BT}{\mu} \int_{-\infty}^\infty  \frac{d^2q}{(2\pi)^2}
				 \frac{e^{-\mu q^2(t-t')/B^t(q)}}{q^2},		 				                  
\eea
for $t>t'$.
Based on this we find
\be \label{eq:MSD}
 C^t(0) - C^t(t) = \frac{k_BT}{\mu} \int_{0}^\infty  \frac{d q}{2\pi}
				        \frac{1-e^{-\mu q^2t/B^t(q)}}{q}
 \equiv \frac{\Delta^t}{2}
.
\ee
where $\Delta^t$ is the transverse contribution to the $\msd$. However, the integral in Eq. (\ref{eq:MSD}) diverges for large wave numbers $q$. To overcome this problem, we introduce the  cutoff $\Lambda$ that  is on the order of the inverse distance between beads in the elastic model. With this cutoff, 
Eq. (\ref{eq:MSD}) reads
\bea
\Delta^t &=& \frac{k_BT}{\mu}\int_{0}^{\Lambda} 
		\frac{dq}{\pi} \frac{1-e^{-\mu q^2t/B^t(q)}}{q}
\eea
The corresponding expression for $\Delta^l$ is the same but with $\mu$ replaced by $\lambda+2\mu$ and $\eta$ by $2\eta$. 

To find the long time dynamics of  $\msd$, we need to find the asymptotic behavior of $\Delta^t$ and $\Delta^l$. To that end, we investigate two cases: $\xi\ne 0$ and $\xi=0$. These correspond to a membrane that is supported by a solid interface on one side and viscous liquid on the other ($\xi\neq0$), and liquid on both sides ($\xi=0$, but then with $\eta\to 2\eta$ to reflect the doubling of the liquid).

\subsection{Case 1: $\xi\ne 0$}
Here we substitute $y=q\sqrt{\mu t/\xi}$ to get
\bea
\Delta^t &=& \frac{k_BT}{\mu}\int_{0}^{\Lambda\sqrt{\mu t/\xi}} 
                \frac{dy}{\pi} \frac{1-e^{-y^2\xi/B^t(y/\sqrt{\mu t/\xi})}}{y}
\eea
We now split the integration into two parts, according to whether $y$ is above or below a parameter $\epsilon$ that we choose such that $1\ll \epsilon \ll {\rm min}\{\sqrt{\mu t \xi}/\eta,\Lambda\sqrt{\mu t/\xi}\}$. For $y<\epsilon$ we can approximate $B^t(y/\sqrt{\mu t/\xi})\sim \xi$ and for $y>\epsilon$ we can ignore the exponential term in the numerator. This gives
\bea
\Delta^t &\sim & \frac{k_BT}{\pi\mu}\left(\int_{0}^{\epsilon}
                dy \frac{1-e^{-y^2}}{y} +
\int_{\epsilon}^{\Lambda\sqrt{\mu t/\xi}}
                dy \frac{1}{y}\right)
\eea
From these integrals we find when $\epsilon\gg 1$ that
\be
\Delta^t\sim\frac{k_BT}{2\pi\mu}\left[\ln\left(\frac{\mu\Lambda^2}{\xi}t\right)+\gamma\right]
\ee
where $\gamma=0.577\dots$ is Euler's constant. Again, to get the longitudinal part  $\Delta^l$ we replace  $\mu$ by $\lambda+2\mu$ in $\Delta^t$. Adding  $\Delta^l$ and $\Delta^t$ in Eq. (\ref{eq:msd_c}) gives the asymptotic behavior
\be
\msd \sim \frac{k_BT(\lambda+3\mu)}{2\pi\mu(\lambda+2\mu) }\ln t \propto \ln t
\ee
Thus, a tracer particle in a supported membrane has a MSD proportional to the logarithm of time.

\subsection{Case 2: $\xi=0$} 
In this case we have (after replacing $\eta$ with $2\eta$)
\bea
\Delta^t &=& \frac{k_BT}{\mu}\int_{0}^{\Lambda}
                \frac{dq}{\pi} \frac{1-e^{-\mu qt/(2\eta)}}{q}
\eea
This integral can be  evaluated directly and has the asymptotic behaviour
\be
\Delta^t\sim\frac{k_BT}{\pi\mu}\left[\ln\left(\frac{\mu\Lambda}{2\eta}t\right)+\gamma\right]
\ee
As before, we get $\Delta^l$ by letting  $\mu\rightarrow\lambda+2\mu$ and doubling the viscosity: $2\eta\rightarrow 4\eta$. For large times, we find that
\be
\msd \sim \frac{k_BT(\lambda+3\mu)}{\pi\mu(\lambda+2\mu) }\ln t \propto \ln t
\ee
Thus, like for a supported membrane ($\xi \neq 0$) the MSD grows logarithmically with time.
Note, however, that the result above  differ from the $\xi\neq 0$ -- case by a factor of $\frac 1 2$.

\section{Summary and Concluding Remarks}

In this paper we show that the mean squared displacement of a tracer particle in a solid membrane move logarithmically slow.  This type of logarithmic time evolution could, in some cases,  serve as a complement to existing models used to interpret single--particle tracking experiments, foremost fractional Brownian motion (FBM) and continuous time random walks (CTRW), that predict power--law dynamics. For example, as an alternative interpretation of molecular dynamics simulations of lipids in lipid membranes \cite{jeon12}, where fitting a power--law yields a very small exponent ($\alpha\approx 0.16$). 

The logarithmic dynamics in our model stems from collective motion of the membrane. Such collective behaviour cannot be described by FBM and CTRW that are single particle models that predict power--law dynamics. However, we point out that there are several single--particle processes different from ours that exhibit logarithmic time evolution. We here mention three examples: 
(i) Sinai diffusion \cite{havlin2002diffusion} where a particle moves in  an energy landscape with power--law distributed energy barriers, 
(ii) extreme value problems such as record statistics \cite{schmittmann1999weather} where the logarithm comes from the increasing difficulty to break new records,
(iii) hitchhiking between cities can lead to logarithmic time evolution for the number of cities visited at time $t$ \cite{lomholt2013microscopic}. See also \cite{lomholt2013microscopic} for a larger list of examples of logarithmic time evolution.

\acknowledgements
 LL acknowledges the Knut and Alice Wallenberg foundation and the Swedish Research Council (VR), grant no. 2012-4526, for financial support. TA is grateful to VR for funding, grant no. 2014-4305. MAL acknowledges funding from Danish Council for Independent Research - Natural Sciences (FNU), grant number 4002-00428B.


\end{document}